# Wavelength conversion of data at gigabit rates via nonlinear optics in an integrated micro-ring resonator

Alessia Pasquazi[1], Raja Ahmad[2], Martin Rochette[2], Michael Lamont[3], Brent E. Little[4], Sai T. Chu[4], Roberto Morandotti[1], and David J. Moss[1,3]

[1] Ultrafast Optical Processing, INRS-EMT, Université du Québec,
1650 Blv. L. Boulet, Varennes, Québec J3X 1S2 Canada
[2] McGill University, Dept. of Electrical and Computer Engineering, Montréal (PQ), H3A 2A7, Canada
[3] CUDOS, School of Physics, University of Sydney, New South Wales 2006, Australia
[4] Infinera Corp. 9020 Junction drive Annapolis, Maryland, 94089. USA

**Abstract** We present the first system penalty measurements for all-optical wavelength conversion in an integrated ring resonator. We achieve wavelength conversion over a range of 27.7nm in the C-band at 2.5 Gb/s by exploiting four wave mixing in a CMOS compatible, high index glass ring resonator at ~22 dBm average pump power, obtaining < 0.3 dB system penalty.

## 1. Introduction

All-optical signal processing is recognized [1, 2] as being fundamental to meet the exponentially growing global bandwidth demand and low energy requirements of ultra-high bit rate communications systems. The possibility of exploiting ultra-fast optical nonlinearities for the realization of critical signal processing functions has been widely explored in the last two decades. Among the nonlinear phenomena of interest, of particular importance is four wave mixing (FWM) via the Kerr nonlinearity ($n_2$), as it can be used to perform frequency conversion, signal reshaping, optical regeneration and other important all-optical functions [2-6]. However, an optimal efficiency of this, and other, nonlinear optical phenomena still requires significant improvement to ultimately achieve commercial deployment. Ring resonators are seen as a key approach to enhancing the nonlinear efficiency [7] by effectively recycling the optical pump power within the resonant cavity. The use of high Q-factor optical resonators has enabled the demonstration of very low power continuous-wave (CW) nonlinear optics based on pure silica microtoroids and microspheres [8-10], photonic crystal nanocavities [11] or integrated micro-rings in high index glass [12,13] and silicon [14]. Similar benefits are expected [7] in terms of operation on optical signals containing high bandwidth data, although the challenge in this case is that the bandwidth of the resonator must be large enough to accommodate all of the spectral components of the optical signal. It is only very recently [15] that the first demonstration of optical signal processing based on nonlinear optics in a resonant cavity has been reported. In particular, wavelength conversion at 10 Gb/s in silicon cascaded ring-resonator optical waveguide (CROW) devices has been [15] achieved. However, full system penalty measurements have not yet been reported in these structures - only in the context of FWM in straight waveguides and nanowires in silicon [16,17] and in chalcogenide glass [18].

In this paper, we present the first full system penalty, or bit error ratio (BER), measurements for wavelength conversion in an integrated ring resonator. We achieve near error-free all optical wavelength conversion at 2.5 Gb/s in the C-band via four wave mixing (FWM) in a high index doped silica glass ring resonator with a Q-factor of 65,000, a free-spectral range (FSR) of 575 GHz and a full-width at half-maximum (FWHM) of 3 GHz. The device is based on a CMOS compatible high index doped silica glass [12,13] that exhibits an effective waveguide nonlinearity ($\gamma = \omega\, n_2 / c\, A_{eff}$, where $A_{eff}$ is the effective mode area, $c$ is the speed of light, and $\omega$ is the pump frequency) 200 times larger than that of standard single mode fibers [7,8]. We achieve < 0.3 dB power penalty at $10^{-9}$ BER for converting a 2.5 Gb/s signal from 1562 nm to 1535nm (a 27.7nm range), using 165 mW of CW pump power. Further, because our waveguides exhibit negligible nonlinear saturation, or absorption, up to extremely high intensities (25 GW/cm$^2$) [19], we observe no saturation in the device performance.

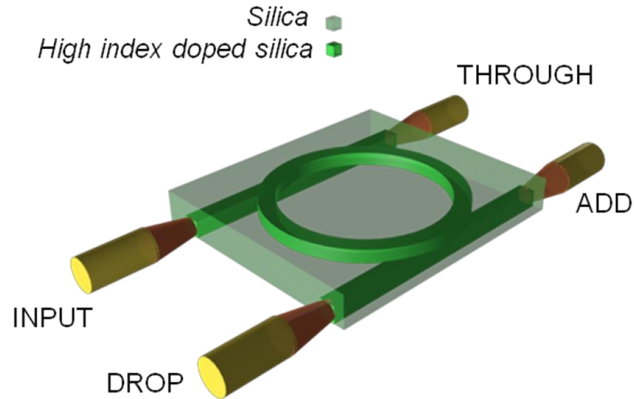

Fig.1 The four-port microring resonator used in the experiments. The trade-name for the high index doped silica glass waveguide core material is Hydex®.

## 2. Experiment

The device under test consists of a four-port micro-ring resonator, 48 μm in radius. The guiding core is high index doped silica glass [12, 13] with a refractive index of 1.7 at 1550 nm, embedded in silica glass and integrated on a silicon wafer. Chemical vapor deposition was used to deposit the glass layers, and the ring resonator was defined with high resolution optical lithography followed by reactive ion etching. The dimensions (and composition) of both the ring and bus are the same, with a rectangular cross section of 1.45 μm x 1.50 μm, yielding a tight modal field confinement due to a large refractive index contrast. We use a vertical coupling scheme between the bus and ring in order to achieve greater control over the gap. Details of the fabrication process are given in [12] and references therein. Chromatic dispersion of this glass waveguide is relatively small with $\beta_2 \cong$ -9 ps$^2$/km at 1550 nm [20]. The ring resonator is coupled to two bus waveguides, as sketched in Fig.1, Each of the four device ports (input, through, add, and drop) is pigtailed, with on-chip mode converters to reduce coupling losses to 1.65 dB. The waveguides support a couple of very weakly bound higher order modes with different symmetry, but they have no impact on the device performance. The coupling scheme ensures that these modes are not excited and we observe no resonances associated with them [12].

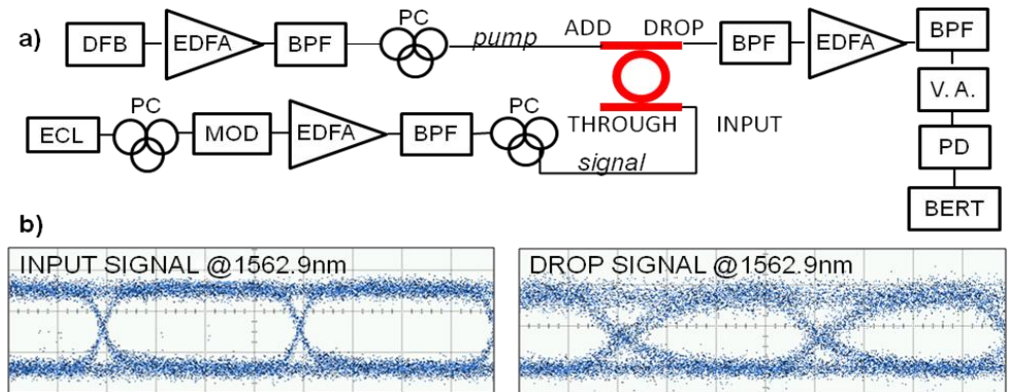

Fig 2 a) Experimental set up for wavelength conversion of a 2.5Gb/s signal (BPF = bandpass filter). DFB= distributed feedback laser, VA=variable attenuator, EDFA=erbium doped fiber amplifier, BPF=band-pass filter, PC=polarization controller, PD=photodiode, MOD=modulator, ECL=edge couple laser. Note, we used an EDFA preamplifier rather than an electronic preamplifier on the photodetector. The variable attenuator was used for the BER measurements. b) Eye diagram of the pseudorandom signal at the input (left), before coupling at the INPUT port, and as collected at the DROP port (right).

Figure 2 shows the experimental setup. The pump was a CW distributed feedback (DFB) tunable laser at λ= 1548.9 nm, amplified by an erbium doped fiber (EDFA). The signal was obtained by modulating a CW tunable external cavity laser (ECL) at λ=1562.9 nm, also amplified with an EDFA, at 2.5 Gb/s, using a pseudorandom bit sequence (PRBS, $2^{15}$-1) in a non return to zero (NRZ) format. We used a bit sequence length of $2^{15}$-1 rather than $2^{31}$-1 in order to improve the stability of the BER measurements, given the relatively low bit rate of 2.5Gb/s. We have verified in other work at higher bit rates that this does not have a significant impact on the results. The pump and signal were filtered using 0.9 nm and 0.8 nm bandwidth filters (BPF), to remove the background amplified spontaneous noise (ASE), and were then coupled into the ADD and INPUT ports, respectively. Two different resonances, offset by three free spectral ranges (FSR) or 1.725 THz (13.8 nm), were excited by the pump and the signal. The effective pump power inside the ring is estimated to be ~10W because of the resonant enhancement of the cavity.

### 3. Results and discussion

The low power signal spectrum had a full width at half maximum of 10pm. The eye diagrams of the signal as coupled into the INPUT port, and collected at the DROP port (output of the sample) are shown in Fig. 2b  Figure 2 shows that the signal eye diagram undergoes only moderate degradation after passing through the ring, and so experiences very little high frequency cut-off after passing through the ring resonance.

The spectra acquired in the presence of a 165 mW pump (coupled power at the ADD bus) for several signal powers are visible in Fig. 3 (top left), measured at the DROP port of the ring resonator. The linear growth in intensity of the generated idler at 1535.2 nm, versus signal power (Fig. 3, bottom left) shows no saturation, and an internal average conversion efficiency of -28.5 dB. The weak peaks visible in the spectrum of the pump at the DROP port are side-modes of the DFB pump laser. The spectrum at the DROP port is not that of the incident pump, but the residual spectrum left after the pump is coupled into the ring, and so the side-modes lying outside the resonance are significantly enhanced relative to the central pump wavelength.

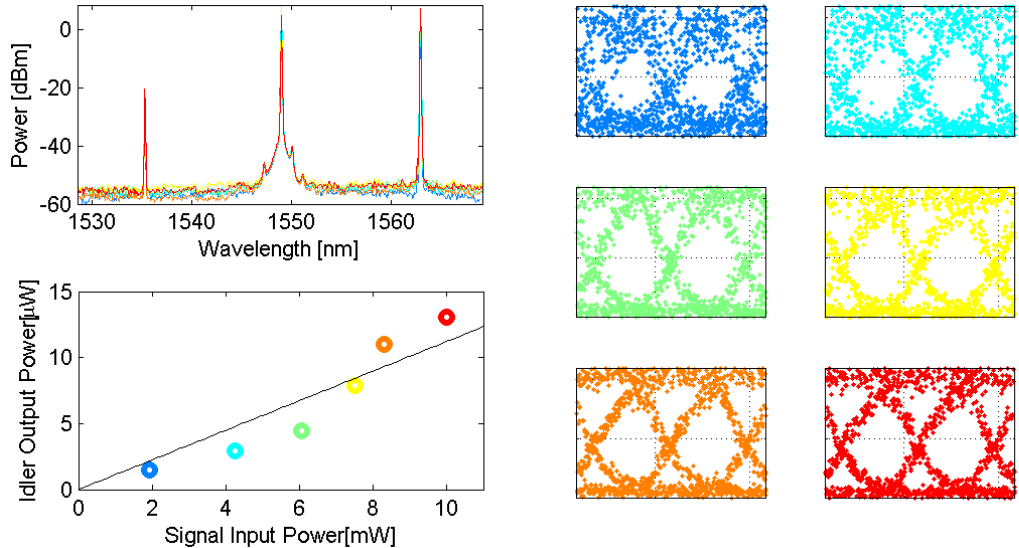

Fig.3 Experimental results obtained with a 165mW pump at increasing signal powers (colors from blue to red ). Top left: OSA spectra collected at the DROP port. Bottom left: idler power versus signal power. Right: Eye diagrams corresponding to the idler in the right panel, acquired with 1350 samples

We did not observe any intensity dependent saturation because of the negligible nonlinear losses in the sample [19]. Note that comparable silicon-on-insulator (SOI) ring resonators typically show saturation for efficiencies ~ -39 dB [14]. A thermo-optical nonlinearity appeared at high pump powers level, however. We observed a slow red shift of the cavity resonances of about 1 nm, consistent with an increase in the refractive index of $10^{-3}$. Thermo-optical shifts can be conveniently employed to finely tune the resonance wavelengths, obtaining a stable working point at the desired wavelength [13, 21-22].

To test the wavelength conversion performance of the ring resonator, the idler was collected at the DROP output, filtered and amplified to measure the eye diagrams with a 65 GHz bandwidth sampling oscilloscope. The BER testing curves and eye diagrams are presented in Fig. 4 for the converted idler and for the signal at the output of the device, for fixed signal and power levels of 10mW and 165mW, respectively. Note that we used an optical preamplifier (EDFA) in place of an electronic preamplifier on the photodiode, which is why the received power levels are comparatively high. Both approaches yield equivalent system penalties. The BER results associated with the idler signal show a remarkably low system power penalty at $10^{-9}$ BER of < 0.3 dB. The idler eye diagram is smoothed slightly with respect to the coupled signal (Fig.2b), indicating that the generated wavelength is not exactly centered at the resonance. This accounts for part of the observed 0.3 dB penalty. Figure 4 indicates that there may be an onset of a slight noise floor just above $10^{-9}$ BER, although this is difficult to conclude definitively since the effect is comparable to the experimental scatter in the data. We do not believe this is a fundamental limitation. The quoted penalty of 0.3dB was obtained by comparing linear fits to the data. The lowest experimental BER we achieve is within a factor of 2 of "error free". Figure 3 shows that the quality of the idler eye diagrams improves significantly with increasing signal power. We would therefore expect the BER penalty to be improved by using higher signal powers, which was limited experimentally in our case to 10mW. This will be reported in future work.

This work demonstrates a practical performance benchmark for ring resonators and a basis for nonlinear all-optical signal processing based on these devices. Following the improvements in spectral efficiency expected from more advanced modulation formats and from the use of lower Q resonators, as well as from the realization of higher order filter designs to improve the bandwidth, we anticipate that the next generation of high-index microring resonators will be capable of operation at bit rates of 40Gb/s and higher. This platform has also demonstrated high efficiency parametric gain [23], supercontinuum generation [24], and many other functions.

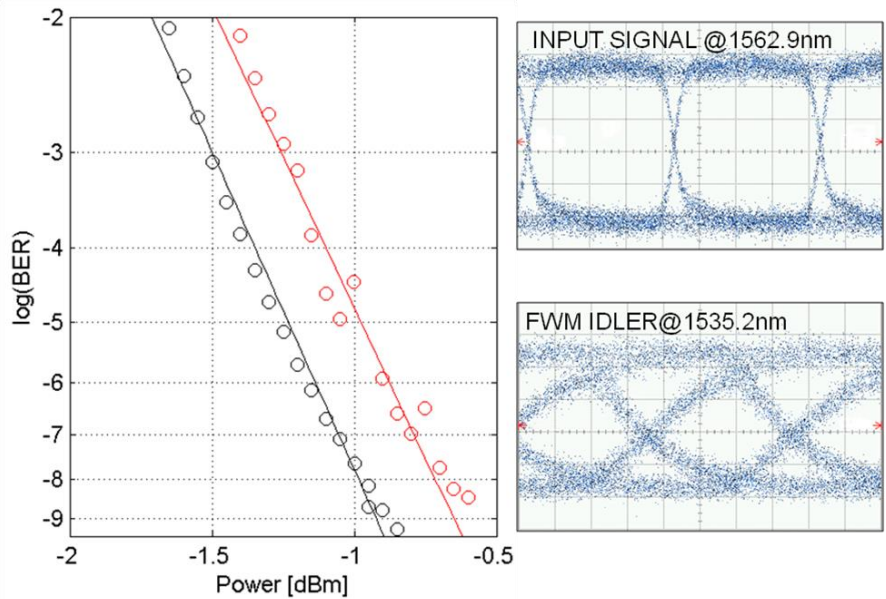

Fig.4 Left: BER measurement for the signal (black line) and the idler (red line) generated by a 10mW signal and a 165mW pump: the power penalty of the two curves is 0.3dBm Right: eye diagram for the signal (top) and the idler (bottom)

## 4. Conclusions

We present the first system penalty measurements of all-optical wavelength conversion in an integrated ring resonator. We achieve wavelength conversion in the C-band at 2.5Gb/s via four wave mixing in a CMOS compatible high index silica glass ring resonator with a Q factor of 65,000. The resulting system penalty of < 0.3dB paves the way for ring resonators to enable low power, low bit error ratio operation for future network telecommunication systems.

## Acknowledgements

We acknowledge financial support of the Natural Sciences and Engineering Research Council of Canada (NSERC), of the FQRNT (Fonds Québécois de la Recherche sur la Nature et les Technologies) as well as of the Australian Research Council (ARC).